\documentclass[runningheads]{apinv}
\usepackage{epsfig,graphics}
\usepackage[T2A]{fontenc}
\usepackage[cp1251]{inputenc}
\usepackage[english]{babel}

\begin{document}

\title{Parameters of 20 newly detected eclipsing binaries from the $Kepler$ database}
\titlerunning{Parameters of 20 eclipsing binaries from the $Kepler$ database}
\author{D. Dimitrov\inst{1}, D. Kjurkchieva\inst{2}, V. Radeva\inst{2}}
\authorrunning{D. Dimitrov, D. Kjurkchieva, V. Radeva}
\tocauthor{D. Dimitrov} \institute{Institute of Astronomy, Bulgarian Academy of Sciences
    \and Department of astronomy, Shumen University\newline
    \email{dinko@astro.bas.bg};
    {d.kyurkchieva@shu-bg.net};
   {veselka.radeva@gmail.com} }
%% \papertype can be "research report", "review", "conference talk", "conference poster", %% \ "lecture at scientific

\papertype{conference talk}

\maketitle

\begin{abstract}
The paper presents a sample of newly detected eclipsing binaries from the public $Kepler$ data.
Orbits and fundamental parameters of 20 unknown eclipsing binaries were determined by
modeling of their photometric data. Most of them are well-detached, high-eccentric binaries. We established
that the target KID8552719 satisfied all widespread criteria for a planetary candidate.
Fitting its light curve we obtained radius
R$_{p}$=0.9 R$_{Nept}$, distance to the host star $a$=42.58 R$\sun$=0.198 AU
and equilibrium temperatute T$_{p}$=489 K. These values imply a Neptune-size object out of the
habitable zone of the host star.
\end{abstract}

\section*{Introduction}

The \emph{Kepler} Mission of NASA is specifically designed (Borucki et al. 2010; Koch et al. 2010; Batalha et al. 2010;
Caldwell et al. 2010; Gilliland et al. 2010) to survey a portion of our region of the Milky Way galaxy to discover
Earth-size planets in/near the habitable zone. The second research area of the \emph{Kepler} mission is
asteroseismology. Moreover some certain objects are observed by ordering of so called "guest-observers".

It was found that one-meter class Schmidt telescope with FoV of 105 square degrees would fulfill the goal of the
\emph{Kepler} mission. This photometer is equipped by 95-megapixel focal plane composed of 42 science CCDs and four
fine guidance sensor CCDs. The CCDs are read out every 6.54 s and
then summed into 29.4 min Long Cadence bins. In addition, up to 512 targets can be observed in Short Cadence mode, at a
59-sec sampling. The observed band spans 423-893 nm, i.e. the \emph{Kepler} magnitudes (Kp) are usually within 0.1 of an R magnitude. The dynamic range is from Kp=7 to Kp=17.

Preliminary counting stars brighter than V=14 from the USNO star catalog, the Cygnus region has been chosen as the
richest area. The Kepler Input Catalog (KIC) of the stars of the observed field (containing 13.2 million targets) was
made preliminary. The photometer sees roughly half million targets. But because it is telemetry-limited, not every
pixel is read out and stored. Only around 190000 pre-selected stars of interest are observed. The selection criteria were determined according to the primary goal of the mission (Batalha et al. 2010).

\emph{Kepler} mission was launched in March 2009. The observations are organized into three-month intervals called
quarters defined by the roll maneuvers the spacecraft executes about its boresight to keep the solar arrays pointed
toward the Sun (Haas et al. 2010). The data packages are processed and analyzed by a Science Pipeline.

2321 planet candidates have already been detected until June 2012, 61 of them are confirmed (Koch et al. 2010, Borucki et al. 2010, Dunham et al. 2010, Latham et al. 2010, etc.).

Another very important achievement of the {\emph{Kepler} mission is discovery of great numbers of variable stars which
light curves are with unprecedented precision. Many eclipsing binaries were discovered and classified automatically on
this rich data base.

Prsa et al. (2011, Paper I) catalogs 1879 eclipsing and ellipsoidal binary systems identified in the first Kepler data
release (Borucki et al. 2010). Slawson et al. (2011, Paper II) provide an updated catalog augmented with the second \emph{Kepler} data release which
increases the baseline nearly to 125 days. As a result 386 new systems have been added while 100 systems have been
removed from the previous catalog. With these changes, the total number of identified eclipsing binary systems in the
\emph{Kepler} field-of-view has increased to 2165, that is 1.4 $\%$ of the Kepler target stars.

The detection of the EBs presented in the foregoing catalogs are based primarily on the initial pipeline calibrated
light curves. Prsa et al. (2011) used automated approach to identify plausible EBs, which entails morphological classification of each light curve's power spectrum.
Blomme (2011) proposed improvement of the method for automate classification.

Answering to the appeal to use the available resources of the {\emph{Kepler} database for additional research we tried
to search for additional eclipsing binaries using the public data from \emph{Kepler} sets Q0, Q1 and Q2. This paper
presents the procedure and results from this work.

\section{Our searching method}

Our searching of eclipsing binaries began in 2011 on the basis of the public data of quarters Q0-Q2 of the {\emph{Kepler} mission. Initially we used the data from the web-based NASA Exoplanet Archive
http://exoplanetarchive.ipac.caltech.edu/\\
applications/ETSS/Kepler/
index.html. Further we used also the access to these data via
ftp://130.167.252.39/pub/kepler/lightcurves/. It should be noted that we considered
only the long cadence data because the short cadence ones were with bigger scatter

The photometric data for each quarter and KID are calibrated for bias (dark level) and converted to fluxes. They are available in two forms: "RAW" and "CORR".

Light curves for all observed stars (processed on a quarterly basis at NASA Ames Research Center) are placed in the
archive. Users of {\emph{Kepler} data might work with both versions of the light curves. Particularly, everyone may use
de-trended data or make de-trending himself.

During 2011 the web-based NASA Exoplanet Archive contained not only stellar parameters (magnitudes, temperatures, etc.) but also the statistical parameters of their light curvesl
"Light curve chi-squared", "Points in light curve beyond 5 $\sigma$" and "Fraction of points in light curve beyond 5 $\sigma$".
We used the values of these statistical parameters to search for eclipsing binaries and managed to pick around thousand EB candidates. Unfortunately, the web-based NASA Exoplanet Archive do not contain yet the foregoing statistical parameters and the considered method for selection of EB candidates is not applicable.

The next step was visual examination of our EB candidates. It revealed that most of them are false positives and only around hundred are true EB.

Besides the foregoing approach we made visual inspection of all light curves of one third of the Kepler data (until folder 0040 from the downloaded ftp data).

As a result of these two ways of searching we gathered candidates that were out of the published catalogs of $Kepler$ eclipsing binaries (Prsa' catalog version 2.0).

Further we checked if some of these candidates were known variables (to the middle of May 2012). For this aim we used:

(i) Lists of Planetary Candidates, False Positives and Eclipsing Binaries from the site
http://archive.stsci.edu/kepler/results.html;

(ii) The AAVSO database of variable stars (International Variable Star Index)

(iii) Databases SIMBAD and ADS.

Several candidates turned out known variables and were removed from our sample.
Thus we made the first version of our list of EB candidates.

In this paper we present part of the newly detected eclipsing stars satisfying the additional condition
to have at least 3 eclipses from different type.

During the visual inspection of the light curves of the \emph{Kepler} database we established that they need additional
preliminary "manual" treatment before detailed analysis because the maximum fluxes as well as the light amplitudes for
many stars differ from one quarter (even subset) to another. Moreover, the "CORR" data are not always properly
de-trended. That is why we decided to use the RAW data of our targets and to made own normalization and de-trending of
the data. For this aim the data from each subset were reproduced by spline function and normalized separately.

At the beginning of 2012 when the quarters Q3-Q6 became free-accessed we used the new data
for more precise determination of the ephemeres of our targets.

The last examination of our targets consisted in checking for possible mis-classified of variability of some target as a result of
contamination of nearby known binary with the same period. This effect is due to the high star density near the
galactic plane and to the fact that the \emph{Kepler} pixels projected onto 4 arcsec on the sky. Thus, for instance, we found that the unknown variable star KID3098197 (detected by us) and the known EB star KID3098194 (Paper I) at distance 4 arcsec have the same period and type of variability (Fig. 1). It is difficulty to say which of
these stars is the true variable because the contamination depends on the relative magnitude of the stars in question
and their separation on the sky. As a result of the examination for contamination effect (into area with radius 60
arcsec around our candidates) several candidates as KID3098197 were excluded from our list.

\begin{figure}[!htb]
\begin{center}
    \centering{\epsfig{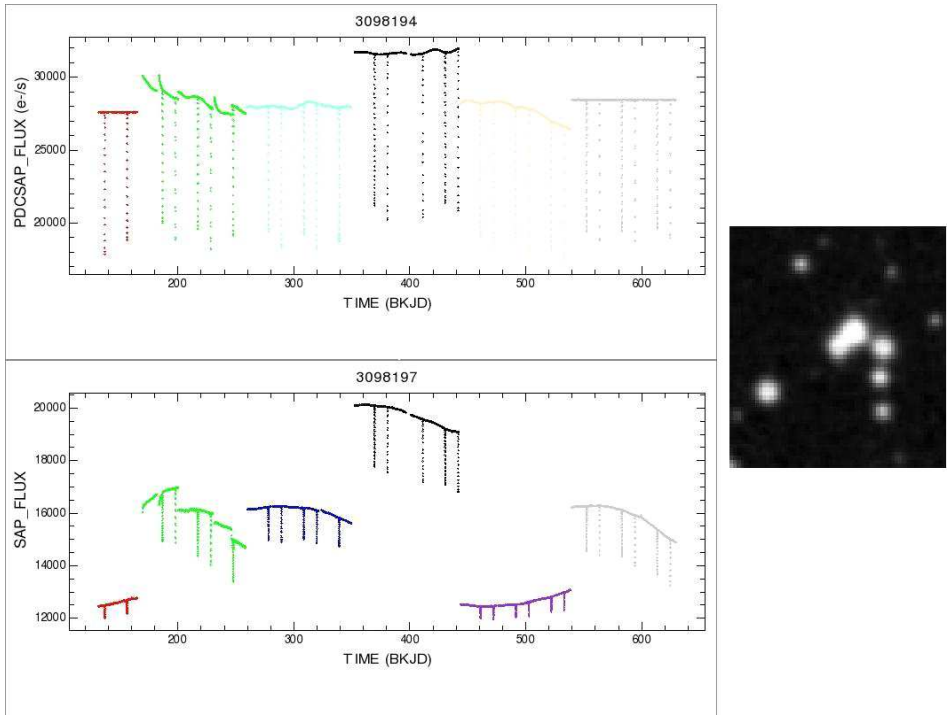}}
    \caption[]{Illustration of contamination effect by the stars KID3098194 and KID3098197}
    \label{1}
  \end{center}
\end{figure}

\section{Light curve modeling}

As a result of our searching method we found around 150 newly detected EB from the \emph{Kepler} database
Q0-Q2. This paper presents 20 of them (Sample 1) which periods and light curves are well determined and allow detailed
analysis. Table 1 presents information for their coordinates and ephemeres while Figure 2 shows fields around them with sizes 1 arcmin.

\begin{table}
\begin{center}
\caption{Coordinates and ephemeres of the targets from Sample 1}
\begin{tabular}{ccccr}
\hline
\\
KID     & RA[2000]       & DEC[2000]      &  BJD0           &   Period  \\
\\
\hline
\\
1995732 & 19h 04m 51.84s & 37d 25m 30.86s & 2455049.45692971 & 77.363032  \\
2307206 & 19h 29m 30.28s & 37d 39m 26.86s & 2455086.53717057 &204.030667  \\
3102000 & 19h 09m 23.87s & 38d 17m 09.74s & 2454997.25556938 & 57.059216  \\
3428468 & 19h 06m 53.61s & 38d 35m 44.38s & 2455032.44996961 &114.908636  \\
3851949 & 19h 27m 36.80s & 38d 55m 33.10s & 2454976.78185172 & 54.776991  \\
4247023 & 19h 07m 23.64s & 39d 18m 17.78s & 2455061.78039575 & 47.189567  \\
4659255 & 19h 32m 11.61s & 39d 43m 31.26s & 2455067.15120620 &198.014033  \\
6182849 & 18h 52m 17.77s & 41d 32m 27.17s & 2454967.70080728 &  6.399107  \\
6233483 & 19h 56m 12.01s & 41d 32m 14.82s & 2454976.85610603 & 15.873552  \\
6312534 & 19h 55m 32.15s & 41d 39m 43.49s & 2454965.94158660 &  3.015550  \\
6525209 & 19h 30m 53.13s & 41d 54m 17.71s & 2455017.42871952 & 75.131944  \\
6933781 & 19h 06m 47.36s & 42d 27m 04.18s & 2455062.75755611 &130.419512  \\
6947666 & 19h 26m 01.46s & 42d 26m 53.20s & 2455071.15247452 & 56.687032  \\
7021177 & 19h 10m 32.89s & 42d 31m 50.95s & 2455017.49723690 & 18.645334  \\
7132542 & 19h 43m 32.45s & 42d 37m 18.77s & 2455067.74039272 & 66.360950  \\
7620844 & 19h 42m 26.49s & 43d 15m 37.51s & 2455070.18823707 & 58.621582  \\
7866921 & 18h 39m 25.46s & 43d 40m 04.44s & 2455107.26924060 & 91.500871  \\
8044608 & 19h 47m 36.97s & 43d 49m 26.00s & 2455067.64037980 &106.175893  \\
8491745 & 19h 22m 02.79s & 44d 33m 21.67s & 2455059.64770915 &111.320439  \\
8552719 & 19h 15m 53.20s & 44d 37m 28.31s & 2455009.27199064 & 88.407500  \\
\end{tabular}
\end{center}
\end{table}

\begin{figure}[!htb]
\begin{center}
    \centering{\epsfig{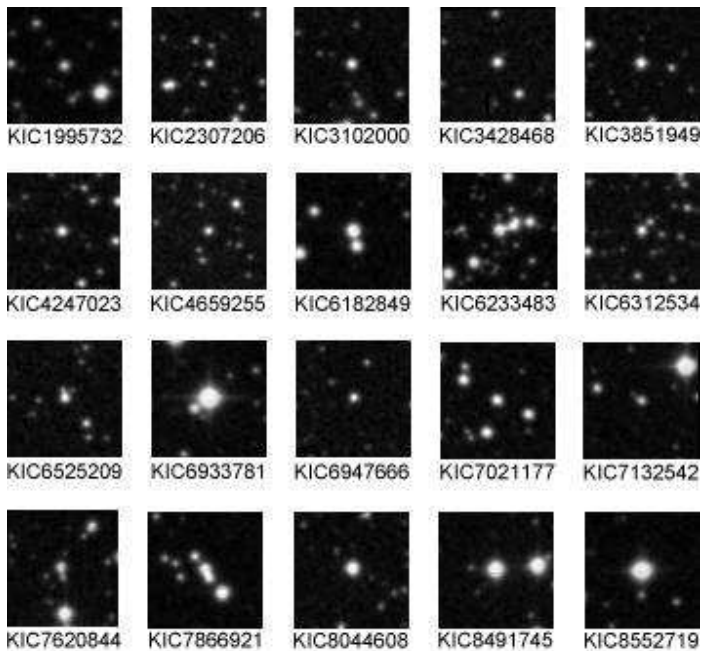}}
    \caption[]{Fields around the targets (at the centers) with size 1 arcmin}
    \label{2}
  \end{center}
\end{figure}

Figures 3-6 exhibit the parts of the phased light curves around the eclipses (the longer out-of-eclipse parts are almost flat).

\begin{figure}[!htb]
\begin{center}
    \centering{\epsfig{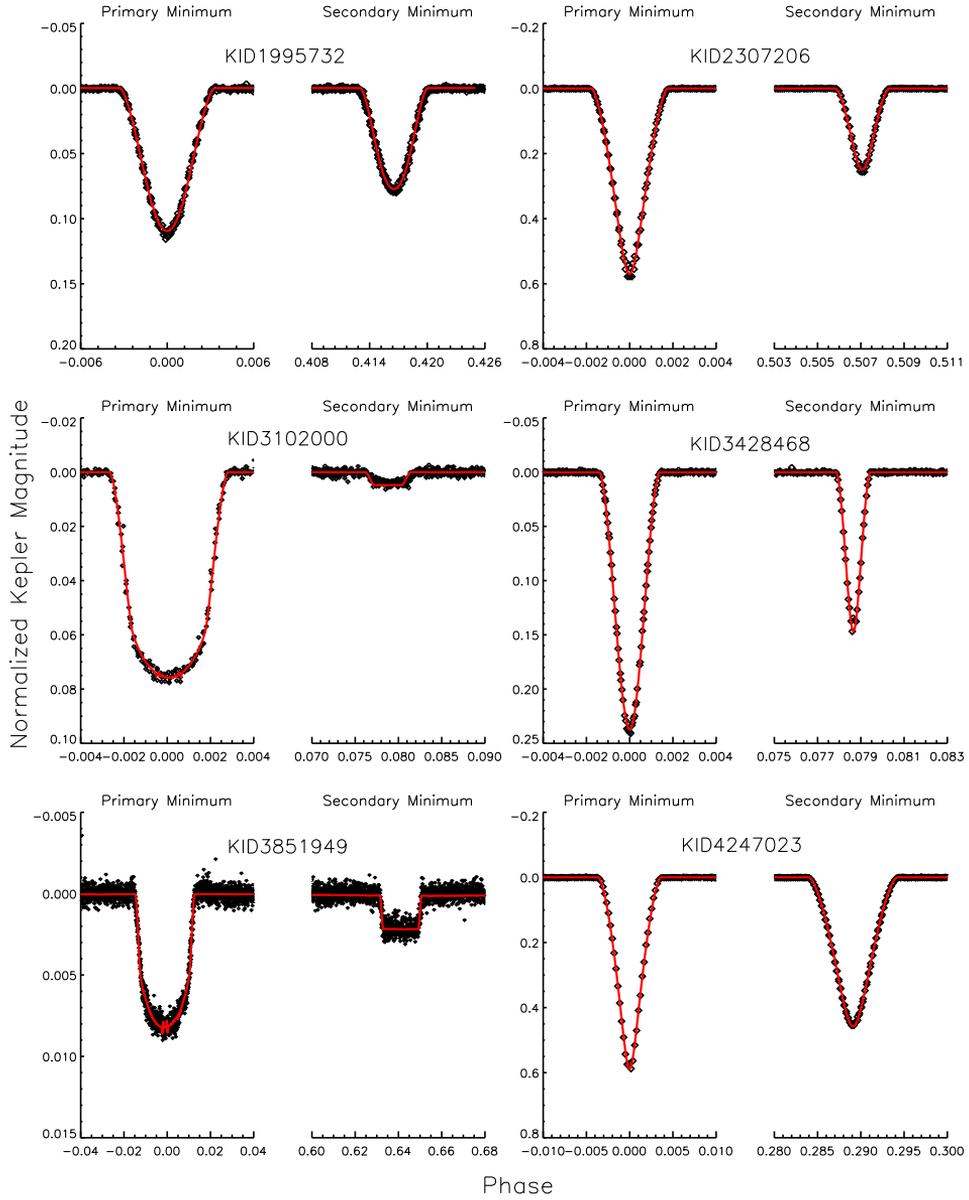}}
    \caption[]{Eclipses of KID1995732, KID2307206, KID3102000, KID3428468, KID3851949, KID4247023 and their fits}
    \label{3}
  \end{center}
\end{figure}

\begin{figure}[!htb]
\begin{center}
    \centering{\epsfig{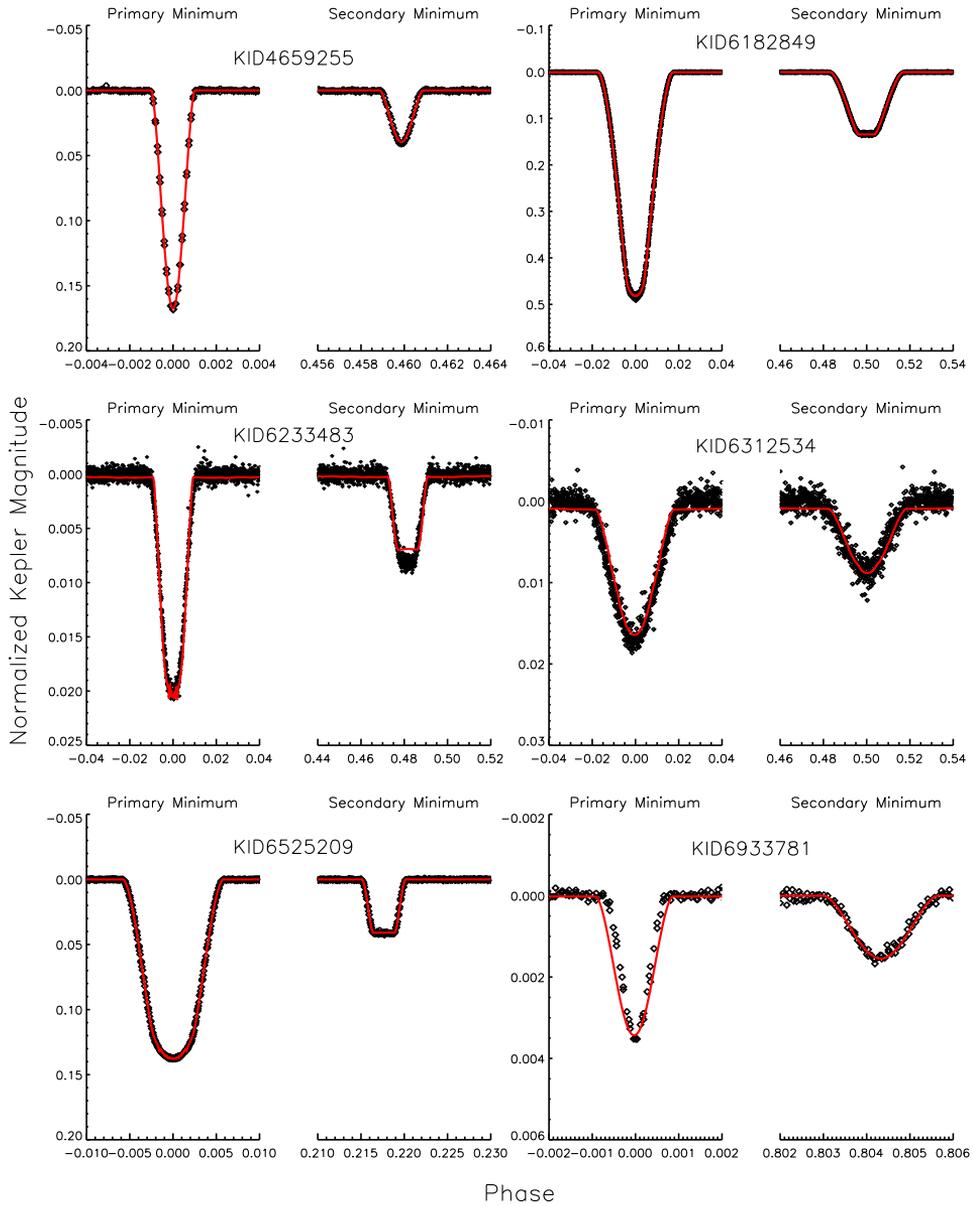}}
    \caption[]{Eclipses of KID4659255, KID6182849, KID6233483, KID6312534, KID6525209, KID6933781 and their fits}
    \label{4}
  \end{center}
\end{figure}

\begin{figure}[!htb]
\begin{center}
    \centering{\epsfig{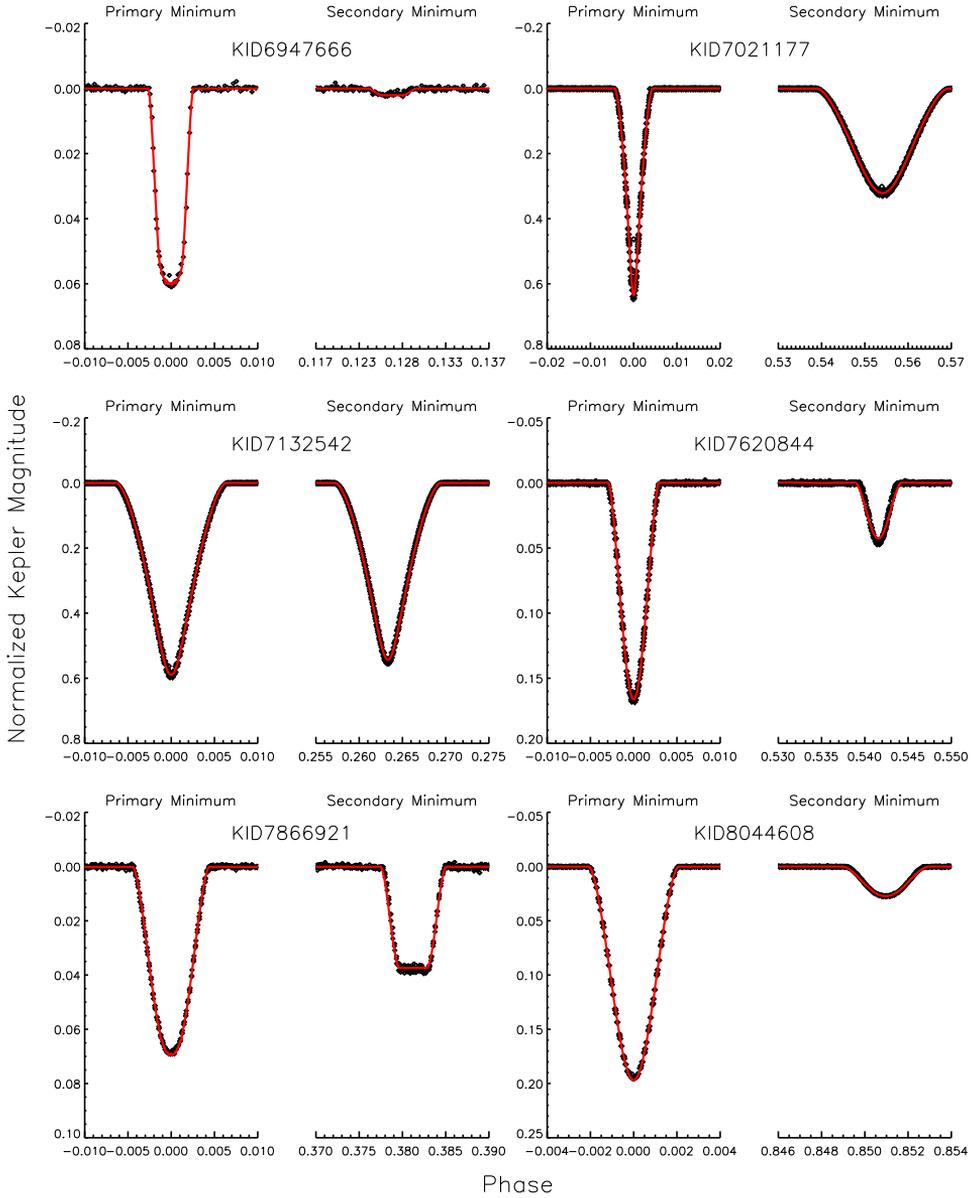}}
    \caption[]{Eclipses of KID6947666, KID7021177, KID7132542, KID7620844, KID7866921, KID8044608 and their fits}
    \label{5}
  \end{center}
\end{figure}

\begin{figure}[!htb]
\begin{center}
    \centering{\epsfig{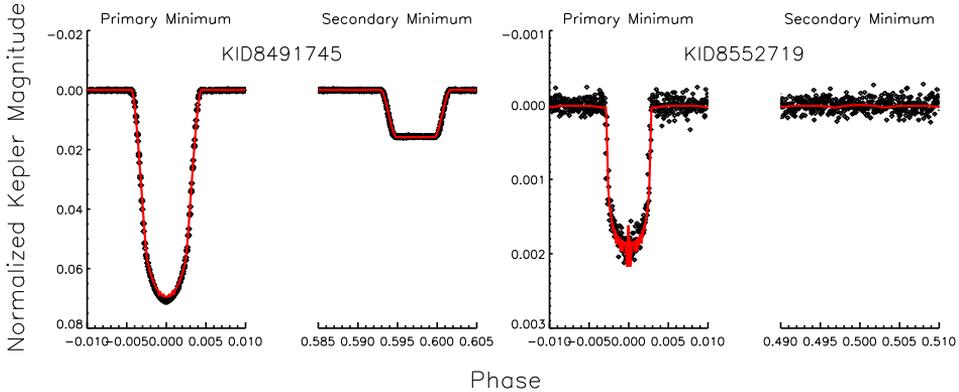}}
    \caption[]{Eclipses of KID8491745, KID8552719 and their fits}
    \label{6}
  \end{center}
\end{figure}

Table 2 summarizes the parameters of the eclipses (phase of the secondary eclipse $\varphi_{2}$, durations $d_{i}$ and
amplitudes $A_{i}$ of the two minima) as well as the orbit parameters $e$ and $\omega$ determined by modeling.

\begin{table}
\begin{center}
\caption{Parameters of the eclipses and the orbit of the targets}
\begin{tabular}{crcrcccr}
\hline
\\
KID & $\varphi_{2}$ & $d_{1}$&$ d_{2}$& $A_{1}$& $A_{2}$ & $e$          & $\omega$        \\
        &        & days   & days   & mag    & mag    &                  & 			\\
\hline
\\
1995732 & 0.4165 & 0.4951 & 0.5415 & 0.1430 & 0.1040 &	0.14252	&	157.08	\\
2307206 & 0.5071 & 0.7141 & 0.4897 & 0.7600 & 0.3410 &	0.17664	&	273.55	\\
3102000 & 0.0790 & 0.3081 & 0.2970 & 0.0950 & 0.0045 &	0.73492	&	181.77	\\
3428468 & 0.0787 & 0.3103 & 0.1724 & 0.3100 & 0.1800 &	0.77405	&	207.65	\\
3851949 & 0.6425 & 1.5007 & 1.0187 & 0.0080 & 0.0020 &	0.29014	&	319.65	\\
4247023 & 0.2891 & 0.4719 & 0.3303 & 0.6570 & 0.4920 &	0.37693	&	151.88	\\
4659255 & 0.4599 & 0.4040 & 0.3723 & 0.1900 & 0.0450 &	0.07572	&	213.66	\\
6182849 & 0.5000 & 0.2240 & 0.2150 & 0.5240 & 0.1388 &	0.01178	&	269.91	\\
6233483 & 0.4813 & 0.3492 & 0.3175 & 0.0132 & 0.0047 &	0.05569	&	238.72	\\
6312534 & 0.5000 & 0.1206 & 0.1206 & 0.0203 & 0.0087 &	0.00115	&	 15.42	\\
6525209 & 0.2176 & 0.8565 & 0.3757 & 0.1480 & 0.0399 &	0.58949	&	224.72	\\
6933781 & 0.8044 & 0.1826 & 0.3391 & 0.0033 & 0.0014 &	0.56295	&	 32.14	\\
6947666 & 0.1260 & 0.2834 & 0.2834 & 0.0628 & 0.0019 &	0.63304	&	175.39	\\
7021177 & 0.5543 & 0.1604 & 0.5482 & 0.7088 & 0.3385 &	0.57877	&	 83.09	\\
7132542 & 0.2632 & 0.8361 & 0.7831 & 0.6767 & 0.6277 &	0.38159	&	182.84	\\
7620844 & 0.5416 & 0.3517 & 0.2579 & 0.3500 & 0.0892 &	0.18617	&	290.21	\\
7866921 & 0.3814 & 0.7869 & 0.6497 & 0.0760 & 0.0430 &	0.22484	&	214.15	\\
8044608 & 0.8510 & 0.4247 & 0.3716 & 0.2019 & 0.0283 &	0.58766	&	356.54	\\
8491745 & 0.5973 & 0.9574 & 0.9462 & 0.0712 & 0.0158 &	0.15338	&	  0.33	\\
8552719 & 0.5000 & 0.7073 & 0.7073 & 0.0020 & 0.0000 &	0.00000	&	180.00	\\
\end{tabular}
\end{center}
\end{table}

The qualitative analysis of the light curves led to the following notes.

(1) The secondary eclipses of KID3102000 and KID6947666 are around 10 times shallower than the primary minimum.

(2) The flat bottoms of the secondary eclipses of KID3851949, KID6182849, KID6233483, KID6525209, KID7866921, KID8491745 imply
occultations.

(3) The light curve of the most targets exhibits different duration of the two eclipse minima and/or the secondary
minimum is shifted from phase 0.5 that are sure evidences of eccentric orbits (Table 2). The selection effect could explain
partially such a result because the eccentric systems can be identified more surely. Another reason for the relative
big number eccentric orbits is the relative large orbital periods of the targets from our Sample 1. It is known
that eccentric orbits are not too infrequent among well-separated systems of long periods.

The quantitative analysis of the targets was made by modeling of their light curves using the software $PHOEBE$ (Prsa $\&$ Zwitter
2005), the values of the mean temperatures $T_{m}$ from the Kepler Input Catalog (Batalha et al. 2010) and the
limb-darkening coefficients derived by Sing (2009) especially to the \emph{Kepler} wide, nonstandard response function
and avoiding the major deficiency present at the limbs of the 1D stellar models.

It is known that the modeling of eccentric binaries requires much time and sometimes is very difficult without any guessed input values
of the fitted parameters. That is why we calculated preliminary values of the eccentricity $e$ and periastron
angle $\omega$ by the analytical formulae of Kopal (1959). Their using as input parameters shortened considerably the
modeling time. Another trick in this direction was the averaging the huge number of data points out of the eclipses
into phase bins.

The fitting parameters of our procedure and their values derived as a result of the light curve solutions are shown in
Table 3 which columns are: mean temperature T$_{m}$; the ratio of the temperatures of the components
$T_{2}/T_{1}$; orbital inclination $i$; photometric mass ratio $q$; relative radii $r_{1}$ and $r_{2}$; third light L3.
The values of the rest fitting parameters (initial epoch BJD0, period $P$; eccentricity $e$ and periastron angle
$\omega$) are given in Tables 1-2. The fits of the minima of some targets are not perfect and may be considered as preliminary ones. The next public data will be used for improving of these models.

\begin{table}
\begin{center}
\caption{Fundamental parameters of the binaries from Sample 1}
\begin{tabular}{ccrcrrrc}
\hline
\\
KID		& T$_{m}$ & $T_{2}/T_{1}$ &	$i$		&	$q$	&	$r_{1}$	&	$r_{2}$	&	L3	\\
\hline
\\
1995732	&	5163	&	0.959	&	89.1715	&	0.450	&	0.01902	&	0.00689	&	0.1264\\
2307206	&	5342	&	0.897	&	89.8847	&	0.787	&	0.00536	&	0.00514	&	0.0037\\
3102000	&	5718	&	0.683	&	89.9889	&	0.218	&	0.02003	&	0.00472	&	0.0000\\
3428468	&	4279	&	0.860	&	89.3941	&	1.830	&	0.00597	&	0.00506	&	0.0000\\
3851949	&	4981	&	0.840	&	88.8893	&	0.102	&	0.07074	&	0.00549	&	0.0000\\
4247023	&	5424	&	0.987	&	89.6098	&	1.003	&	0.01492	&	0.01459	&	0.0000\\
4659255	&	5153	&	0.784	&	89.6561	&	0.992	&	0.00425	&	0.00444	&	0.0000\\
6182849	&	5893	&	0.816	&	89.2096	&	1.254	&	0.06713	&	0.04041	&	0.0003\\
6233483	&	5942	&	0.808	&	86.8016	&	1.000	&	0.06943	&	0.00991	&	0.0000\\
6312534	&	4897	&	0.894	&	77.3068	&	0.300	&	0.01480	&	0.09830	&	0.1307\\
6525209	&	5207	&	0.851	&	89.5055	&	0.450	&	0.02113	&	0.00690	&	0.0000\\
6933781	&	5502	&	1.045	&	89.6300	&	1.000	&	0.00362	&	0.00579	&	0.9850\\
6947666	&	5777	&	0.629	&	88.8576	&	0.080	&	0.02041	&	0.00456	&	0.0000\\
7021177	&	5684	&	0.985	&	89.2692	&	1.000	&	0.02896	&	0.02442	&	0.0000\\
7132542	&	5389	&	0.989	&	89.5995	&	1.260	&	0.01929	&	0.02335	&	0.0000\\
7620844	&	5056	&	0.741	&	89.1450	&	1.000	&	0.01070	&	0.01265	&	0.0000\\
7866921	&	5019	&	0.904	&	89.0497	&	0.202	&	0.02444	&	0.00629	&	0.0000\\
8044608	&	5788	&	0.718	&	89.1088	&	0.350	&	0.01207	&	0.00666	&	0.0000\\
8491745	&	4839	&	0.837	&	89.4061	&	0.236	&	0.02405	&	0.00564	&	0.0000\\
8552719	&	5114	&	0.196	&	89.4149	&	0.038	&	0.02009	&	0.00080	&	0.0000\\
\end{tabular}
\end{center}
\end{table}

\section{Analysis and comments on the results}

The analysis of the obtained parameters led to several important conclusions.

(1) Almost all orbital inclinations, excluding the shortest-period target KID6312534, are near 90$^{0}$ that is normal
for long-period eclipsing binaries.

(2) The relative stellar radii are less 0.07 that means that the systems are well-detached.

(3) The temperatures of stellar components are in the range 3500-6000 K. The primaries are the hotter components. The temperatures of the secondaries are smaller with less 1500 K than those of the primaries. The only exclusion is the target KID8552719 which secondary' temperature is of planetary' order.

(4) The values of the photometric mass ratio $q$ are in the range 0.20-1.83. It shoult be noted that the $q$ values are precisely determined only for the targets with total eclipses.

(5) Although the presence of eccentric systems in our sample was expected (Table 2) it was surprising to find 7 of the
20 targets to have $e > $0.5 (the biggest value is $e$=0.77). It should be noted that the final values of the
eccentricity and periastron angle differ from the preliminary ones by only several $\%$.

The examination of the Eclipsing Binary with Eccentric Orbits Catalog (Bulut $\&$ Demircan 2007) revealed that 22 targets have $e >$ 0.3 and only 2 of them are with $e >$ 0.5. In fact, this catalog
containing information for 124 systems with eccentric orbits, at present is the largest catalogue of eccentric binaries compiled from the literature including three large sources Hipparcos catalogue, the atlas of Kreiner, Kim $\&$ Nha (2001) and the ninth catalogue of spectroscopic binary orbits (Pourbaix et al. 2004).

The examination of the catalogs from PaperI and PaperII revealed that 1376 are eccentric systems: 7 with $e >$ 0.6; 34 with $e >$0.5; 184 with $e >$ 0.3; 799 with $e >$ 0.1 (the biggest value is $e$=0.64).

It should be noted that the most eccentric binary 41 Dra has a period of 1247 d and $e$=0.97. The last value is
determined by radial velocity curve (Tokovinin et al. 2003).

The presented statistics exhibit the relative big frequency of high-eccentric targets in our Sample 1. We attributed
this result mainly to the huge duration of uninterrupted observations made by Kepler.

(6) Besides the eclipses with orbital period of P$_{orb}$=3.0156 d the target KID6312534 exhibits a variability with
amplitudes around 4.5 times smaller than the depths of the primary eclipses and
a period of around P$_{2}$=3.005 days which values slightly change from one subset to another.
It is reasonable to attribute the second variability to spots and the period P$_{2}$ to the stellar rotation.
This means that KID6312534 is synchroneous that is normal for a such orbital period. The small
changes of P$_{2}$ may due to the differential rotation of the spotted star. It should be noted that there were several subsets in which the spot variability lacks at all. This fact may due to some artificial effect. 

After the submission of our manuscript we were informed that KID6312534 is known eclipsing binary which light curve has been 
investigated by Coughlin et al. (2011). The reason of this discrepancy is that this target presents in the Prsa' catalog version
1.0 but misses in the newer version 2.0 we checked.  

(7) The modeling of KID6933781 was possible only with huge value L3=0.98. This result implies that the modeled light
variability does not belong to the observed star but rather to a nearby star which light enters into the aperture around
the observed target. The map of KID6933781 (Fig. 3) revealed central bright star and faint star at distance of 10
arcsec or 2.5 pix. Taking into account that the aperture of the $Kepler$ photometry is with radius 5-6 pix it is
reasonable to suppose that the eclipsing star in this case is the faint star.

(8) Besides the binary period of 18.6 days the system KID7021177 exhibits light changes with a period of P=6.355932
days (Fig. 8) which light curve means possible rotational variability (caused by spots, etc.).

\begin{figure}[!htb]
\begin{center}
    \centering{\epsfig{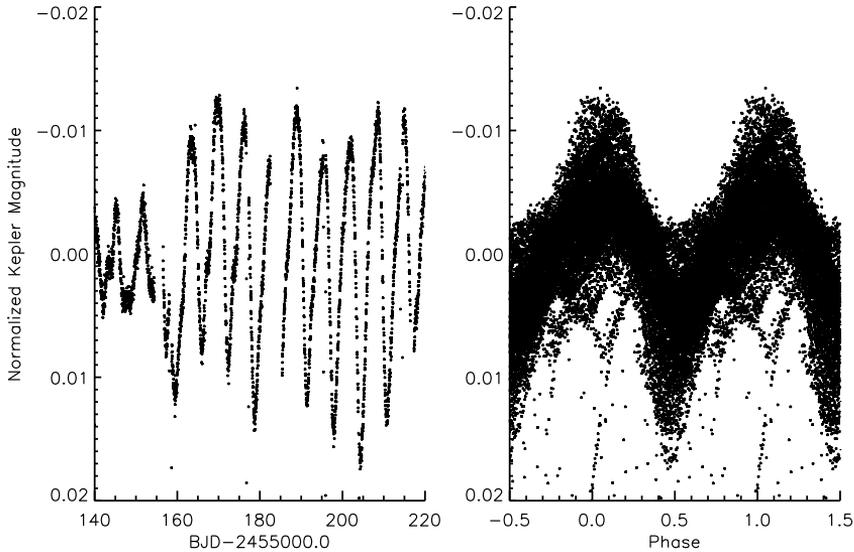}}
    \caption[]{The additional (rotation) variability of KID7021177 and its folded curve}
    \label{8}
  \end{center}
\end{figure}

(9) The depth of the primary eclipse of KID8552719 is only 0.002 mag and there
is not any indication of a secondary eclipse (Fig. 6). Hence, this target reveals
behavior of an exoplanet candidate. The shape of the light minimum also
resembles transit event (not V-shape).

\section{The exoplanet candidate KID8552719}

The planet detection passes through data validation (DV)
and follow-up observations (FOP). DV performs false-positive
elimination from diagnostics that can be pulled out of the \emph{Kepler}
data itself (Batalha et al. 2010), while the FOP relies on additional observations such
as moderate and high-precision spectroscopic radial velocities.

If the modeled light curve of given exoplanet candidate
leads to radius $<$ 2R$_{J}$ it is assigned a Kepler Object of Interest (KOI) number.
But only those candidates that pass the DV tests are submitted to
the follow-up observers for spectroscopic vetting, confirmation,
and characterization. Candidates with a companion radius larger than 2R$_{J}$ are likely associated
with late M dwarfs and are not considered as high-priority planetary candidates.

The goal of the DV tests is to eliminate the false positives caused by either
grazing or diluted eclipsing binaries.

\begin{figure}[!htb]
\begin{center}
    \centering{\epsfig{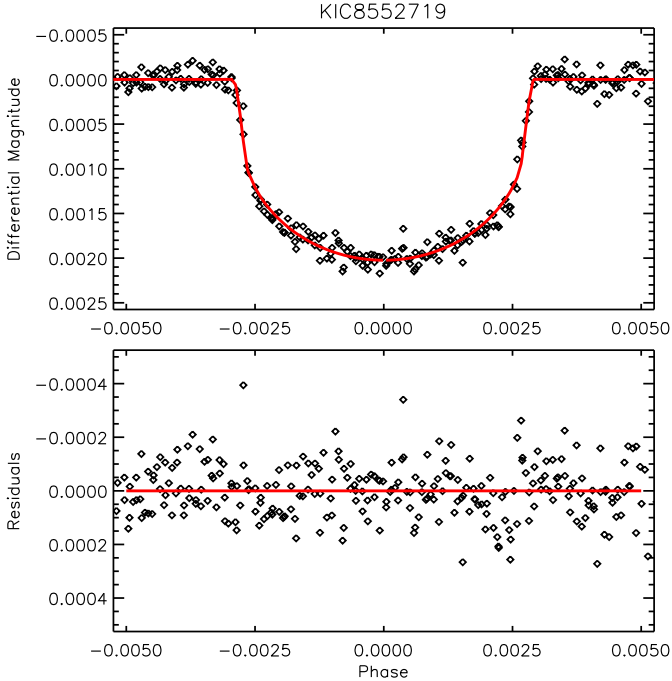}}
    \caption[]{Model of KID8552719 by our code}
    \label{9}
  \end{center}
\end{figure}

We used own code (Kjurkchieva $\&$ Dimitrov, 2012)
to model the light curve of KID8552719.
For this aim we retrieved the star parameters from the Kepler Input Catalog:
T$_{s}$=5114 K (Sp K0 V); R$_{s}$=0.822 R$\sun$; $\lg g_{s}$=4.56.
We calculated the stellar mass M$_{s}$=0.94 M$\sun$. With these
initial values a transit fit was computed (Fig. 9).
In fact, the fitted parameters were R$_{s}$/a, R$_{p}$/R$_{s}$, orbital inclination $i$,
limb-darkening coefficient $u$, initial epoch T0 and period $P$.

As a result of modeling we obtained: orbital inclination $i$=89$^{0}$.53;
orbital radius $a$=42.58 R$\sun$=0.19757 AU; R$_{p}$=0.0321 R$\sun$=0.315 R$_{J}$=0.9 R$_{Nept}$.
Thus, KID8852719 turned out almost Neptune-sized object and passed
the first criterion for exoplanet candidate.

The next criterion (Batalha et al. 2010) required modeling of the odd
and even numbered transits independently.
A significant difference in the companion radii (10 $\sigma$ difference) for the odd and even transit
implies that this system is likely a diluted EB at twice the orbital period.
Our candidate did not reveal considerable difference of the odd and even transits
and passed this criterion.

A false positive can be identified by presence of secondary events at phase 0.5
with depth greater than 2$\sigma$. Our examination for phase 0.5 did not reveal
such a event. In any case, in order to escape cases as KID310200 and KID694766 (see Figs. 3 and 4)
with shallow secondary events far away from phase 0.5, we examined the whole
light curve of KID8552719 but did not find such a secondary event.
Hence, our candidate passed this test as well.

Assuming the star and companion behave as black bodies
we estimated the equilibrium temperature T$_{eq}$ of the secondary component
of KID8552719 by the formula (Batalha et al. 2012)
\begin{equation}
T_{eq} = T_{s}(R_{s}/2a)^{1/2}[f (1 - A_{B})]^{1/4}
\end{equation}
where $A_{B}$ is a Bond albedo and $f$ is a a proxy for atmospheric thermal circulation.
Using the values $A_{B}$=0.1 (for highly irradiated planets, Rowe
et al. 2006) and $f$=1 for efficient heat distribution to the
nightside of planet we obtained a rough estimate for the dayside
temperature of the secondary T$_{eq}$=489 K. This value is nearly that of dayside Mercury
and is suitable to a planet. It should be noted that it is higher than the temperature limit
of the habitable zone of the host star.

% hot Jupiter planet candidates (Jupiter-size candidates with orbital periods near 3 days)
% warm Jupiters (with slightly larger orbits) and hot Neptune-size candidates

The next test for false positives is checking of possible shift of the photocenter
just at the transit (Batalha et al. 2010).
If there are transit-like features in both the row and column centroid residuals and they
are highly correlated with the transit features in the flux time series this is evidence of
a diluted EB, i.e. false positive. In our case there were not any shifts of the photocenter of KID8552719
during the transits.

Another diagnostics to confirm the identity of the transiting
(or eclipsing) object is to examine the spatial flux distributions and
to find if the out-of-transit photocenter is centered on the target star.
For this test one should compare the out-of-transit image with
the difference image. If their brightest pixels are different then the event is
possible a false positive. The last step is to make a time series analysis
of the brightest pixels that confirms which is the source of the transit
variability.

The planetary candidate KID8552719 passed also all photocenter motion diagnostics
(the centroid time series, the difference image, and the pixel flux time series) and
it may be included in the next version of the planetary candidate catalog.
We estimated that if the planet mass is near that of Neptune (as its radius)
then the mass ratio of the system should be around 10000 and the effect on the radial
velocity of the host star would be around 1 m/s.

During the final stage of preparation of this manuscript we found a paper of
Huang, Bakos and Hartman appeared in the astro-ph.EP on May 29 2012 (Paper III).
These authors reported new 16 planetary candidates based on searching for in the Q1-Q6 public data.
One of them is KID8552719.

Huang, Bakos $\&$ Hartman (2012) used the formalism of Mandel $\&$ Agol (2002) and
fitted the parameters R$_{p}$/R$_{s}$, the square of the impact parameter  $b$=$a \cos i$/ R$_{s}$
and the inverse of half duration $\zeta$/R$_{s}$ = 2/T$_{dur}$ that is related to a/R$_{s}$ and the impact parameter $b$.
	 	
The comparison of our results with those in Paper III revealed that the values of the planetary radius coincide within 1 $\%$ while the values of the rest geometric parameters
(orbital inclination and semi-major axis) differ by 2 $\%$. We attributed the last differences  to the different methods of normalization of the $Kepler$ data.

There is not estimation of the planet temperature in Paper III.

\section{Conclusion}

Applying own searching method for eclipsing binaries to the public
$Kepler$ database Q0-Q2 we detected candidates that were out of the published catalogs of the  $Kepler$ eclipsing stars.

This paper presents the results of our light curve solutions of 20 newly detected eclipsing binaries and the values of their orbital and fundamental parameters. Most of them turned out binaries
with high-eccentric orbits and consequently present appropriate targets for study of apsidal motion and the internal structure of the component stars.

Some of our discoveries have interesting peculiarities. We confirmed the conclusion of Huang, Bakos $\&$ Hartman (2012) that KID8552719 satisfied the criteria for a planetary candidate. The estimations for its physical parameters revealed the Neptune-size object out of the habitable zone of the host star.

Hence, the softwares do are indispensable for an automatic processing and analysis of a huge data-sets of the space missions as \emph{Kepler} but the obtained results in this
paper revealed that it is deserved to examine such data-sets "manually/visually" in order to compensate the omission of
some interesting objects and events, available in the database but not registered by the software. At first glance this
time-consuming work seems boring and requires patience and attention. But it is a true pleasure to see the wonderful
light curves of the \emph{Kepler} mission!

\section*{Acknowledgements}
This study is supported by funds of the project DO 02-362 of the Bulgarian Ministry of education and science. is
partially supported by projects DO 02-362, DO 02-85 and DDVU 02/40-2010 of the Bulgarian National Science Fund.

This research has made use of the NASA Exoplanet Archive, which is operated by the California Institute of Technology,
under contract with the National Aeronautics and Space Administration under the Exoplanet Exploration Program, the SIMBAD 
database, operated at CDS, Strasbourg, France, and NASA's Astrophysics
Data System Abstract Service.

\end{document}